\documentclass[onecolumn,preprint,aps,prb,raggedbottom,superscriptaddress]{revtex4-1}

	\usepackage{lmodern}			
	\usepackage[latin1]{inputenc}
	\usepackage[T1]{fontenc}

	\usepackage{graphicx}     % figures
	\usepackage{amsmath}      % equations
	\usepackage{tabularx}		  % tables
	\usepackage{xcolor}			  % colored text

	\usepackage{epstopdf}			% include eps-files
	\usepackage{hyperref}   	% use for hypertext links
	\usepackage[all]{hypcap} 	% Link jumps to the top of a graphic/table instead of jumping to its caption

	\usepackage{blindtext}		% add blind text
	\usepackage{ulem}					% underline text or cancel it out

	\hypersetup{pdfstartview={XYZ null null 0.777}} % Zoom = 77.7%

	\hypersetup{pdfpagemode=UseNone}

	\newcommand{\esubA}{\epsilon_\text{sub,1}}
	\newcommand{\esubB}{\epsilon_\text{sub,2}}
	\newcommand{\esub}{\epsilon_\text{sub}}
  \newcommand{\eaa}{\epsilon_\text{11}}

  \newcommand{\kt}{k_\text{B}T}
  \newcommand{\EES}{E^\text{ES}}
  \newcommand{\mum}{\mu\text{m}}

\begin{document}

\linespread{1.2}

\title{Thin Film Growth of Phase-Separating Phthalocyanine-Fullerene Blends: A Combined Experimental and Computational Study}

\author{Berthold~Reisz}
\author{Eelco~Empting}
\author{Matthias~Zwadlo}
\author{Martin~Hodas}
\author{Giuliano~Duva}
\affiliation{Institute for Applied Physics, University of T\"ubingen, Auf der Morgenstelle 10, 72076 T\"ubingen, Germany}

\author{Valentina~Belova}
\affiliation{Institute for Applied Physics, University of T\"ubingen, Auf der Morgenstelle 10, 72076 T\"ubingen, Germany}
\affiliation{European Synchrotron Radiation Facility, 71, avenue des Martyrs CS 402200, 38043 Grenoble Cedex 9, France}

\author{Clemens~Zeiser}
\author{Jan~Hagenlocher}
\affiliation{Institute for Applied Physics, University of T\"ubingen, Auf der Morgenstelle 10, 72076 T\"ubingen, Germany}

\author{Santanu~Maiti}
\affiliation{Institute for Applied Physics, University of T\"ubingen, Auf der Morgenstelle 10, 72076 T\"ubingen, Germany}
\affiliation{J\"ulich Centre of Neutron Science (JCNS-1), Forschungszentrum J\"ulich GmbH, 52425 J\"ulich, Germany}

\author{Alexander~Hinderhofer}
\author{Alexander~Gerlach}
\author{Martin~Oettel}
\affiliation{Institute for Applied Physics, University of T\"ubingen, Auf der Morgenstelle 10, 72076 T\"ubingen, Germany}

\author{Frank~Schreiber}
\affiliation{Institute for Applied Physics, University of T\"ubingen, Auf der Morgenstelle 10, 72076 T\"ubingen, Germany} 
\affiliation{Center for Light-Matter Interaction, Sensors \& Analytics LISA+, University of T\"ubingen, Auf der Morgenstelle 15, 72076 T\"ubingen, Germany}

\begin{abstract}
\noindent
Blended organic thin films
have been studied during the last decades 
due to their applicability in organic solar cells.
Although their optical and electronic features have been examined intensively,
there is still lack of detailed knowledge about
their growth processes and resulting morphologies,
which play a key role for the efficiency of optoelectronic devices 
such as organic solar cells.
In this study, 
pure and blended thin films 
of copper phthalocyanine (CuPc) and the Buckminster fullerene (C60)
were grown by vacuum deposition onto a native silicon oxide substrate
at two different substrate temperatures, 310\,K and 400\,K.
The evolution of roughness was followed
by \textit{in-situ} real-time X-ray reflectivity.
Crystal orientation, island densities and morphology 
were examined after the growth 
by X-ray diffraction experiments
and microscopy techniques.
The formation of a smooth wetting layer 
followed by rapid roughening 
was found in pure CuPc thin films,
whereas C60 shows a fast formation of distinct islands
at a very early stage of growth.
The growth of needle-like CuPc crystals 
loosing their alignment with the substrate
was identified in co-deposited thin films.
Furthermore, 
the data demonstrates that
structural features become larger and more pronounced 
and that the island density decreases by a factor of four
when going from 310\,K to 400\,K.
Finally, 
the key parameters roughness and island density
were well reproduced on a smaller scale 
by kinetic Monte-Carlo simulations
of a generic, binary lattice model with simple nearest-neighbor interaction energies.
A weak molecule-substrate interaction caused a fast island formation
and a weak interaction between molecules of different species
was able to reproduce the observed phase separation.
The introduction of different same-species and cross-species Ehrlich-Schw\"obel barriers
for inter-layer hopping was necessary to reproduce the roughness evolution in the blend
and showed the growth of CuPc crystals
on top of the thin film in agreement with the experiment.
\end{abstract}

\maketitle

\section{Introduction}
\label{sec:intro}

\noindent % Organic semiconductors in general
Organic semiconductors are a class of materials, 
which offer a wide range of possibilities for basic research and technical applications.
For this purpose, thin films of small-molecule organic semiconductors 
have been studied during the last decades
by means of evaporation in vacuum,
which provides both layer thickness control and a clean environment 
\cite{Forrest_2004_Nature,Witte_2004_JMaterRes,Schreiber_2004_PhysStatSola}.
Binary blends of small organic molecules 
serving as donor-acceptor systems in photovoltaic cells 
represent one of the technical applications.
In particularly, 
many small organic molecules co-evaporated
with the well known Buckminster fullerene C60 
exhibit phase separation 
\cite{Hinderhofer_2012_ChemPhysChem}.
It was shown that 
the degree of phase separation 
and the resulting thin film architecture
play a key role for the solar cell efficiency 
due to the diffusion and dissociation of excitons 
at donor-acceptor interfaces
\cite{Opitz_2010_IEEEJSelTopQuant,Hormann_2014_JPhysChemC,Broch_2018_NatCommun,Anger_2012_JChemPhys}.
There are many studies on the optical properties and solar cell efficiencies,
but the evolution of the underlying thin film architectures 
by self-assembly phenomena is not yet fully understood.
Prior studies already attempted to assign structural features 
to pure domains in phase separated blends 
containing the Buckminster fullerene (C60) 
together with other organic compounds such as
sexithiophene (6T) \cite{Veenstra_1997_SynthMet,Lorch_2016_JApplCrystallogr},
diindenoperylene (DIP) \cite{Banerjee_2013_PhysRevLett,Lorch_2017_JChemPhys},
pentacene (PEN) \cite{Salzmann_2008_JApplPhys}
and copper phthalocyanine (CuPc) \cite{Heutz_2004_SolEnergyMaterSolCells,Sullivan_2004_ApplPhysLett},
but the desired assignment remained unclear.
Only for PEN, it was possible to conclude from the width of X-ray reflectivity peaks
that it forms islands with a height exceeding the nominal film thickness.
%
% CuPc-C60
%
The present study shows needle-like CuPc-crystals protruding from blended CuPc-C60 thin films.
Furthermore,
it investigates the growth 
of pure and blended CuPc-C60 thin films,
both experimentally and by simulations.
The experiments examine the growth
up to a film thickness of approximately 20\,nm 
in real-time and \textit{in situ} during the growth
at two different substrate temperatures, 310\,K and 400\,K,
as well as \textit{in} and \textit{ex situ} after the growth.
The simulations are performed within a generic, two-species lattice model
considering structure formation by particle diffusion
as it was studied for atomic thin films in the past
\cite{Krug_2000_PhysRevB,Michely_2004_book}.
Qualitative agreement 
between simulations and experiments has been achieved,
which gives evidence for the usefulness of such simple models 
to describe actual experiments.
%
% theoretical considerations
%
From the general perspective of structure formation in thin films, 
the CuPc-C60 system is a model system for phase separation. 
Film height and time-dependent phase ordering enter as new aspects 
compared to single-component film growth.
It can be expected 
that phase ordering also influences island distribution and roughness.
General considerations on large length scales indicate 
that also phase-separating binary mixtures exhibit Kardar-Parisi-Zhang (KPZ) scaling
\cite{Kardar_1986_PhysRevLett,Kardar_2000_PhysicaA},
but little is known on smaller length scales. 
Simulations on micro- to mesoscopic length scales are not abundant.
On the one hand, 
there are a few studies on molecular dynamics taking all atoms into account.
%there are a few all-atom studies with Molecular Dynamics which
Those studies attempt to model faithfully the growth process of a specific molecule,
but are generically limited by the small number of particles 
and the necessity to consider much higher deposition rates 
for computational reasons.
Examples are pentacene (PEN) growth 
on C60 \cite{Muccioli_2011_AdvMater} 
or silica \cite{Roscioni_2018_JPhysChemLett} 
and 6T monolayer growth on SiO$_\text{x}$
\cite{Chiodini_2020_ProgOrgCoat}.
Typically, one would use simpler, generic lattice models 
to bridge the microscopic and the mesoscopic scale. 
In the literature, 
such models have been used mainly to discuss epitaxial growth of pure thin films, 
i.e. consisting of only one particle species, 
either for submonolayers \cite{Einax_2013_RevModPhys},
or the 3D growth of mounds \cite{Smilauer_1995_PhysRevB,Siegert_1996_PhysRevE,Leal_2011_JStatMech,Assis_2015_JStatMech}.
Early extensions to binary systems can be found in Refs.\,\cite{Landau_1999_ComputPhysCommun,Tao_2008_Physica}.
Here, 
we will apply a binary model on a cubic lattice to the CuPc-C60 system 
and focus on roughness and island densities as observables to compare with the experiments. 
An important question is whether general trends in these observables
as a function of composition and temperature 
can be explained by the lattice model with its generic parameters.   
%
% paper organization
%
This paper is organized as follows. 
In Secs.~\ref{sec:exp} and \ref{sec:sim}, 
more details about the methods used in experiment and simulation are given. 
Section~\ref{sec:results} presents results 
for which a meaningful comparison between experiments and simulations is possible: 
film roughness, island densities and real-space images. 
Finally, Sec.~\ref{sec:summary} gives a summary and conclusions.

\section{Experiments}
\label{sec:exp}

\begin{figure}[htbp]
		\centering
		\includegraphics[width=0.7\columnwidth]{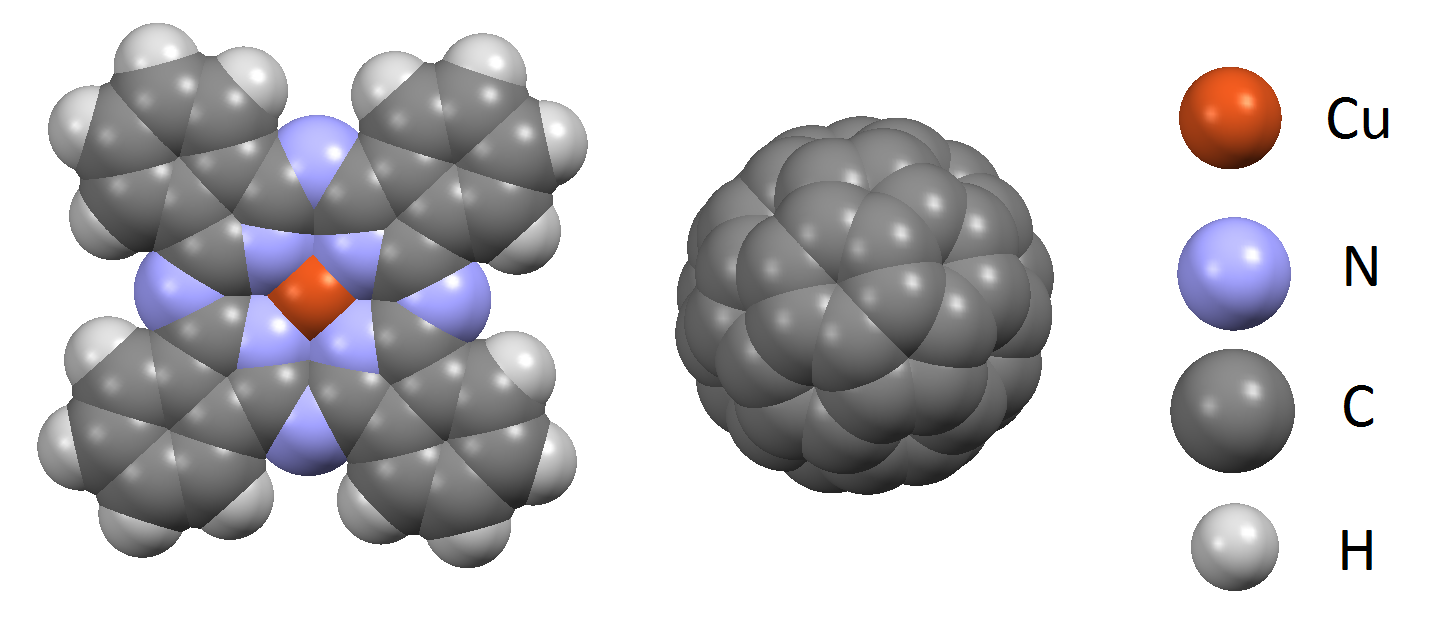}
		\caption{Left: Copper Phthalocyanine (Cu\,N$_{8}$\,C$_{32}$\,H$_{16}$, abbrev. CuPc). Right: Buckminster Fullerene (C$_{60}$, abbrev. C60)}
		\label{fig:molecules}
\end{figure}

\noindent
\textbf{Preparation of organic thin films:} 
Copper phthalocyanine (CuPc) and the Buckminster fullerene (C60), 
were purchased from Sigma Aldrich (purity 99.9\,\% by gradient sublimation), see Fig.~\ref{fig:molecules}.
Native silicon oxide substrates were cleaned by acetone and then by isopropanol
in an ultrasonic bath for 5 minutes each.
Afterwards, 
the substrates were installed inside a vacuum chamber 
for organic molecular beam deposition
and heated up to 500\,K for 10\,hours
until a vacuum in the pressure range 
of $10^{-9}$\,mbar was achieved.
Pure and blended thin films 
containing CuPc and C60
were deposited onto the substrates
at two different substrate temperatures, 310\,K and 400\,K.
The blends were prepared by simultaneous evaporation of CuPc and C60
at a molar mixing ratio of 1:1.
The evaporation temperatures of the individual effusion cells were calibrated in advance 
such that a total deposition rate of 2.0\,{\AA}$/$min was maintained for 100\,min
for both, pure and blended thin films.

% the chemical structure of CuPc and C60

\noindent
\textbf{\textit{In-situ} experiments:}
For the purpose of \textit{in-situ} X-ray diffraction experiments during and after the growth,
a portable ultrahigh vacuum chamber for organic molecular beam deposition 
equipped with a 360$^\circ$-beryllium window
was used \cite{Ritley_2001_RevSciInstrum,Krause_2004_EurophysLett,Krause_2004_SurfSci}.
The \textit{in-situ} real-time and post-growth X-ray reflectivity scans (XRR), 
as well as the \textit{in-situ} post-growth grazing incidence X-ray diffraction scans (GIXD)
were carried out at the material science beamline for surface analysis MS-X04SA of the Swiss Light Source 
using synchrotron radiation with an energy of 12.7\,keV
\cite{Willmott_2013_JSynchrotronRadiat}.
The grazing angle of incidence in GIXD was chosen close to the total reflection edge of silicon
such that the X-ray beam probed the samples throughout the entire film thickness.

\begin{figure*}[htbp]
\includegraphics[width=\textwidth]{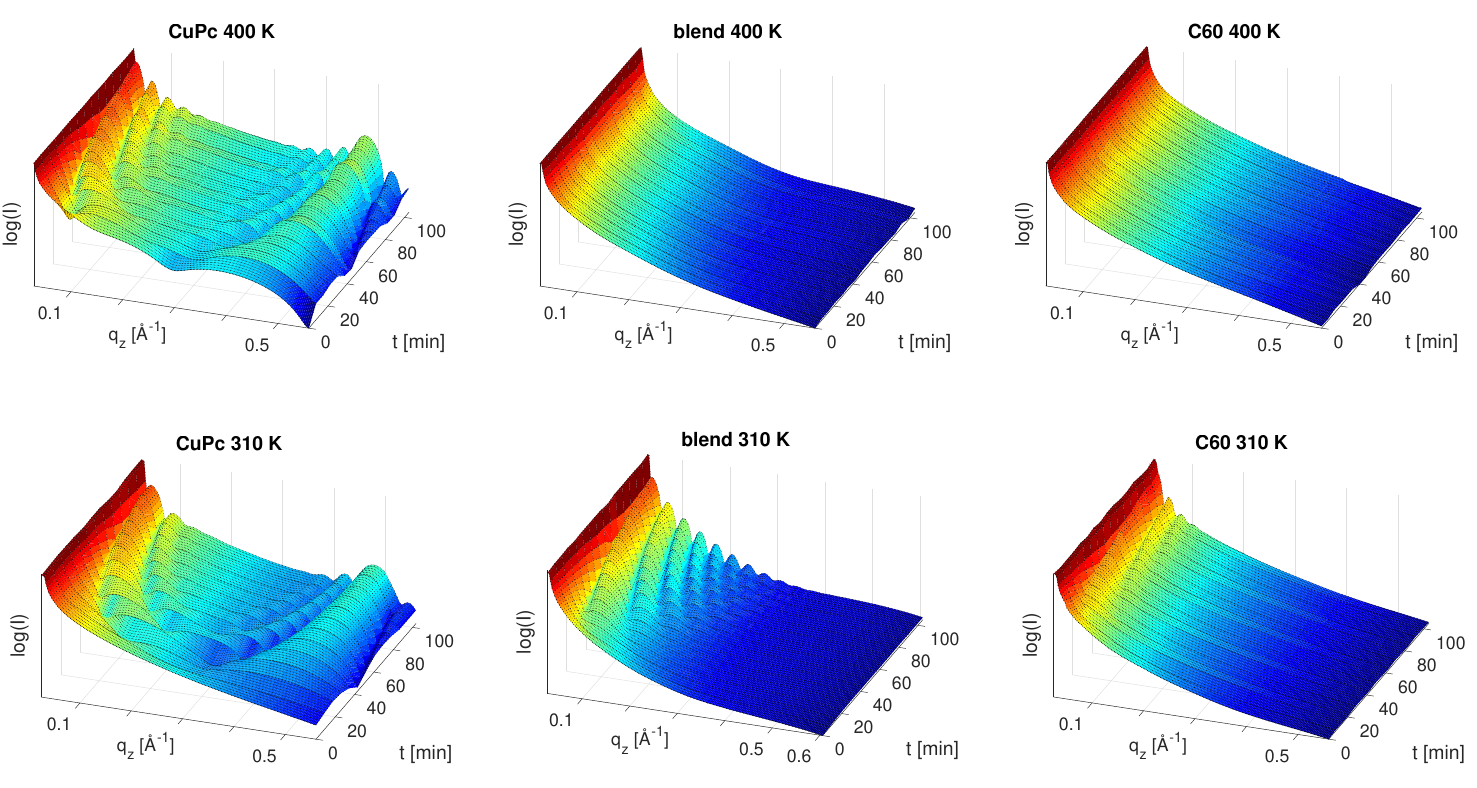}
\caption{X-ray reflectivity (XRR) of 
		pure CuPc (left), pure C60 (right) and a 1:1-blend of both molecules (middle) 
		measured \textit{in situ} and in real-time during the growth
		at two different substrate temperatures: 310\,K (lower row) and 400\,K (upper row).
		The profiles of pure C60 and the 1:1-blend grown at 400\,K exhibit no Kiessig oscillations,
		which means that those two films are already rough from the beginning of growth.}
\label{fig:realtimeXRR}
\end{figure*}

\noindent
\textbf{Real-time XRR:}
The measured intensities $I(q,t)$ of all real-time XRR scans
($q$ is the momentum transfer and $t$ the time)
are shown in Fig.~\ref{fig:realtimeXRR}.
Previous studies investigated the evolution of roughness
by recording the reflected X-ray intensities at specific $q_z$-values
during the growth of pure films.
Those $q_z$-values are the anti-Bragg point $^1/_2\cdot$$q_\text{Bragg}$ 
and $^2/_3$, $^3/_4$, $^4/_5$ and $^5/_6$ of $q_\text{Bragg}$
\cite{Kowarik_2007_ThinSolidFilms,Kowarik_2009_EuropPhysJSpecialTopics,
Lorch_2017_JChemPhys,Durr_2003_PhysRevLett,Zhang_2007_SurfSci,
Bommel_2014_NatComms,Bommel_2015_PhysStatusSolidiRRL,
Yang_2015_SciRep}.
Due to the absence of a clear Bragg peak in the XRR-scans of pure C60 and the blended thin films
there are no anti-Bragg points to be observed.
Instead, full XRR-scans up to $q_z \approx 0.6$\,{\AA}$^{-1}$ were carried out
and the evolution of the root mean square roughness $\sigma$ 
was determined from the damping of Kiessig oscillations
in the low $q_z$-range.
The software GenX, which is based on the Parratt formalism, 
\cite{Bjorck_2007_JApplCrystallogr}
was applied for fitting the Kiessig oscillations 
up to 0.15\,{\AA}$^{-1}$ for pure C60 and up to 0.20\,{\AA}$^{-1}$ for pure CuPc and the blend.
For simplicity, 
the thin films were assumed to be homogeneous with a gradual decrease of the electron density 
in z-direction, which simulates the surface roughness.
A thin void layer was inserted in the model between substrate and thin film
in order to account for a possible depletion of charges at the film-substrate interface.
The XRR scans of the pure C60 and the blended thin film grown at 400\,K
exhibit no Kiessig oscillations due to a pronounced roughness
throughout the entire film growth.
Hence, the evolution of roughness was determined only for the thin films grown at 310\,K.

\begin{figure*}[htbp]
	\centering
		\includegraphics[width=\textwidth]{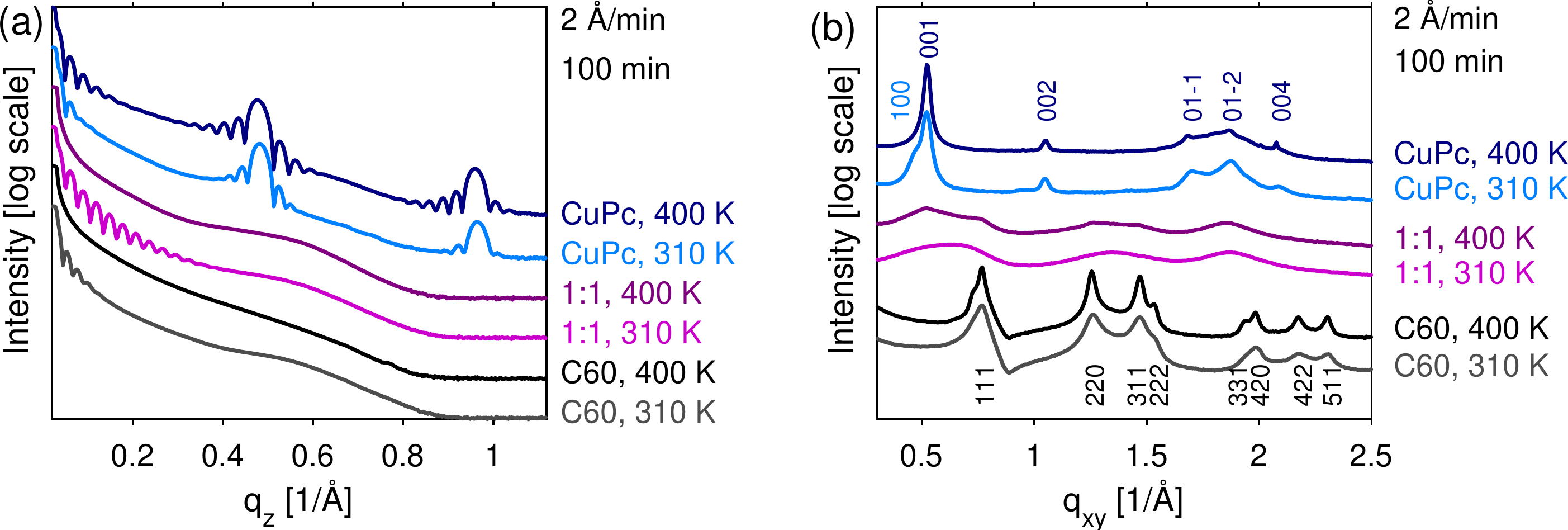}
	\caption{(a) X-ray reflectivity scans (XRR): 
	The influence of mixing and substrate temperature on the roughness is clearly recognizable 
	from the number of visible Kiessig oscillations in the low $q_z$-range. 
	The vanishing of CuPc Bragg peaks in the blends indicates 
	that the presence of C60 hinders the spatial alignment of CuPc crystals with the substrate surface.
	(b) Grazing incidence X-ray diffraction (GIXD): 
	The presence of CuPc and C60 peaks in the profiles of the blended thin films 
	proves their phase separation. 
	The peak broadening	indicates 
	that the lateral sizes of coherently scattering domains became smaller in the blends.}
	\label{fig:XRR_GIXD}
\end{figure*}

\noindent
\textbf{Post-growth XRR:}
Figure~\ref{fig:XRR_GIXD}\,(a) shows the \textit{in-situ} post-growth X-ray reflectivity scans (XRR).
The roughness $\sigma$ 
of each thin film was determined 
from the damping of Kiessig oscillations in the low $q_z$-range 
as described above.
%
%Where applicable, 
The lattice-spacing in out-of-plane direction 
was determined for CuPc from the $q_z$-values of Bragg peaks
and was around 13\,{\AA} at both substrate temperatures.

\noindent
\textbf{Post-growth GIXD:}
Figure~\ref{fig:XRR_GIXD}\,(b) shows the \textit{in-situ} post-growth grazing incidence X-ray diffraction scans (GIXD) 
after transformation into the reciprocal $q_{xy}$-space.
Crystal structures were determined from $q_{xy}$-values 
and the corresponding Bragg peaks were indexed
accordingly to the literature
\cite{Hoshino_2003_ActaCryst,Ashida_1966_BullChemSocJpn,David_1991_Nature,Boer_1994_ChemPhysLett,Hinderhofer_2013_JPhysChemC}.
Occasionally, several ($hkl$)-reflections contribute to the same Bragg peak.
For those Bragg peaks containing multiple ($hkl$)-triplets,
only one triplet is indicated for clarity.
The lateral size of coherently scattering domains $d_{coh}$
was estimated from the GIXD diffraction patterns
by fitting the full width at half of maximum (FWHM) 
of the (001)-peak for CuPc and the (111)-peak for C60
due to the Scherrer Formula $d_{coh} \approx 2\pi/\Delta q_{\text{FWHM}}$
\cite{Patterson_1939_PhysRev}.

\noindent
\textbf{Reciprocal space maps:}
A second set of samples was prepared under the same conditions for \textit{ex-situ} characterization.
Reciprocal space maps of those samples were acquired
at the beamline ID03 of the European Synchrotron Radiation Facility (ESRF)
using synchrotron radiation with an energy of 24.0\,keV.
The grazing angle of incidence was half of the total reflection angle of silicon
such that the measurement was more surface sensitive.
The reciprocal space maps are shown in the supporting information (Fig.~\textcolor{blue}{S1})
and served to determine the crystal orientation \cite{Supporting_Info}.

\noindent
\textbf{AFM, SEM and HIM  images:}
Real-space images of this second set of samples were obtained by \textit{ex-situ} atomic force microscopy (AFM)
using a JPK Nanowizard II instrument operating in tapping mode under ambient conditions.
The islands were automatically counted for the determination of island densities.
The post-growth roughness $\sigma$
was evaluated by $\sigma=\sqrt{ \langle h^2 \rangle - \langle h \rangle^2 }$ 
and compared to the root mean square roughness $\sigma$ determined from post-growth XRR.
$h\equiv h(x,y)$ is the height of the film 
at the lateral position $(x,y)$ in the AFM images.
The data was complemented by scanning electron microscopy (SEM)
at different magnifications
using an XL30-device from Philips at a beam energy of 20\,keV
and by helium ion microscopy (HIM) 
using an ORION Nanofab-device from Zeiss at an acceleration voltage of 30\,kV.
Selected SEM images of the blended thin films are compiled in the the supporting information (Fig.~\textcolor{blue}{S2})
\cite{Supporting_Info}
Selected AFM- and HIM-images 
in comparison to simulated height maps and 3D-snapshots of the simulated growth
will be presented and discussed in Sec.~\ref{sec:Real_Space_Images} below.

% real-time XRR (3d-plots)

% postgrowth XRR and GIXD (2 A/min, 310 K and 400 K)

\noindent
\textbf{Experimental evidence of phase separation:}
The expected phase separation is confirmed by the presence of both CuPc and C60 Bragg peaks
in the GIXD profiles of blended thin films, see Fig.~\ref{fig:XRR_GIXD}~(b).
The peak widths are significantly broadened compared to the GIXD peaks of pure films
and the corresponding size of coherently scattering domains is less than 4\,nm in the blends.
This structure was referred to as nano-crystalline
in previous studies on phthalocyanine-fullerene blends
\cite{Schunemann_2012_PhysRevB}.
For comparison,
the lateral size of coherently scattering domains in the pure films 
increases from 21\,nm to 31\,nm for CuPc
and from 13\,nm to 22\,nm for C60
when raising the temperature from 310\,K to 400\,K.
We see that not only mixing influences the domain sizes, 
but also changing the substrate temperature.
Larger domains are expected 
at elevated substrate temperatures
due to a faster molecular diffusion.
Note that pure CuPc and pure C60 domains in blended thin films
may consist of several nano-crystals with different orientations,
which implies that the pure domains
might be significantly larger
than the size of coherently scattering crystals.
Finally, no new Bragg peaks appear in the blends and hence no new crystal structures are formed.
Although the simulations abstain from all peculiarities 
associated with molecular shapes and crystal structures,
we would like to discuss them briefly
before introducing the simulation methods.
The orientation of crystals was derived from the reciprocal space maps, 
see Fig.~\textcolor{blue}{S1} in the supporting information \cite{Supporting_Info}.
The C60 crystals have no preferential orientation with respect to the substrate surface,
which leads to concentric rings in the reciprocal space
and to the absence of Bragg peaks in the XRR profiles,
see Fig.~\ref{fig:XRR_GIXD}~(a).
The random orientation of C60 crystals on native silicon oxide
and the formation of islands at a very early stage of growth
(as it will be discussed in Sec.~\ref{sec:results})
is in contrast to the growth of C60 on mica
\cite{Bommel_2014_NatComms,Bommel_2015_PhysStatusSolidiRRL}.
It seems that the molecule-substrate interaction plays a crucial role in thin film growth
and was therefore considered as one of the main parameters in the simulations.
Dissimilar to C60,
the CuPc crystals prefer to grow in columns along the substrate surface in a 2D-powder,
which was confirmed by indexing the distinct peaks in the reciprocal space maps of pure CuPc thin films.
Such a 2D-powder consists of randomly oriented crystals in two dimensions,
but is well ordered in the direction perpendicular to the substrate surface,
which can be seen by the well pronounced Bragg peaks in the XRR  profiles of pure CuPc,
see Fig.~\ref{fig:XRR_GIXD}~(a).
These findings are in good agreement with previous studies of vacuum deposited CuPc on different substrates
\cite{Karasek_1952_JAmChemSoc,Suito_1962_Nature,Nonaka_1995_ThinSolidFilms,Reisz_2020_JApplCryst}.
Finally, the reciprocal space map of the blended thin film grown at 400\,K
exhibits a weak diffraction ring corresponding to the CuPc ($100$)-reflection,
which indicates that the CuPc crystals loose their alignment with the substrate 
and are now randomly oriented in three dimensions.
This result is in good agreement with the missing Bragg peaks in the XRR profiles of blends
and it is corroborated by the needle-like crystals protruding from the blended thin films 
(see HIM images in Sec.~\ref{sec:Real_Space_Images}).
Obviously, the needle-like features can be assigned to CuPc crystals due to their columnar growth.
The following section introduces the simulations.
It turned out that it is sufficient to deduce the simulations to a simplified cubic lattice model
considering mainly interaction energies and diffusion rates
in order to reproduce and explain the evolution of roughness and the processes of phase separation.

\section{Simulations}
\label{sec:sim}

\begin{figure}[tbp]
		\centering
		\includegraphics[width=\columnwidth]{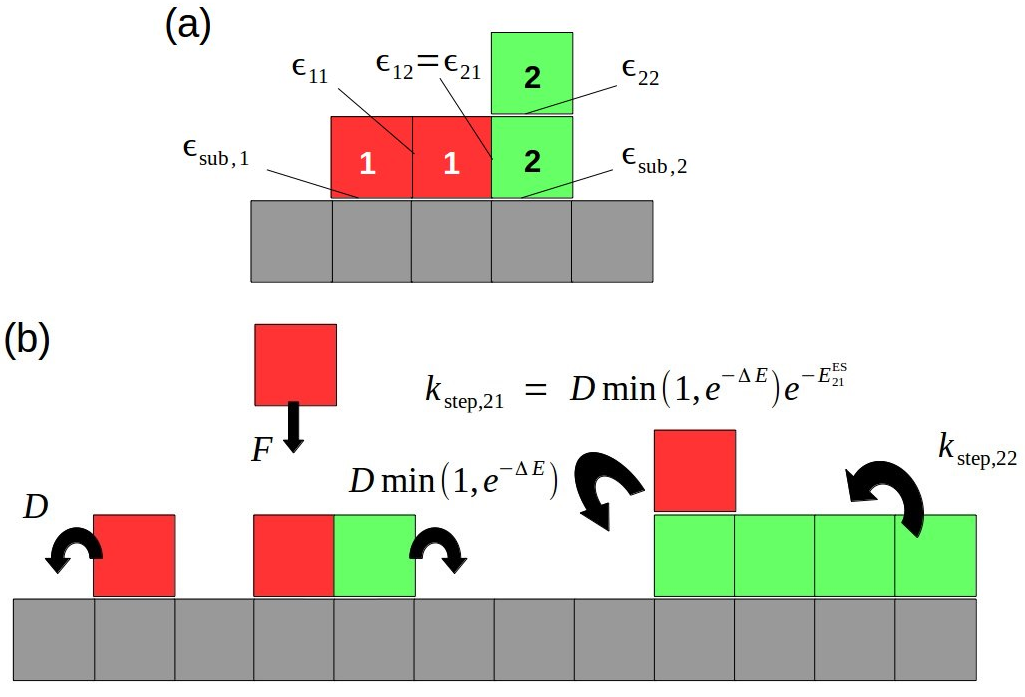}
		\caption{(a) Interaction energies in the lattice model and (b) 
    allowed lattice moves with their rates.}
		\label{fig:sim_defs}
\end{figure}

\noindent
We employ a simple film growth model 
using a binary lattice gas (species 1 and 2) on a cubic lattice.
Nearest-neighbor particles interact with energies $\epsilon_{ij}$ $(i,j=\{\text{1,2}\})$ 
and particles in the first layer have interactions with the substrate ($x$--$y$ plane) 
given by $\esubA$ or $\esubB$ (all energies in units of $\kt$). 
Deposition on top of the film or the bare substrate 
at random substrate plane coordinates is controlled by a rate $F$
(particles per unit time and lattice site). 
Diffusion respects the solid-on-solid (SOS) condition
\cite{Siegert_1996_PhysRevE,Smilauer_1995_PhysRevB}:
Only the particles (species $i$) in the top layer 
are allowed to diffuse to a lateral next-neighbor site 
with rate $k = D \cdot \;\text{min}(1,\exp( -\Delta E))$,
where $D = \kt\,\gamma$ is a free diffusion constant 
and $\gamma\propto e^{-E_\text{D}/(\kt)}$ 
is a surface mobility with Arrhenius-like temperature dependence
featuring a (dimensionful) diffusion barrier $E_\text{D}$.
$\Delta E$ is the energy difference between final and initial state.
In such a diffusion step, 
particles of species $i$ may also ascend 
(moving on top of a particle of species $j$) 
or descend one layer
(moving down from a particle of species $j$) 
in which case the rate is multiplied with
$\exp(-\EES_{ij})$ and $\EES_{ij}$ is an Ehrlich-Schw\"obel (ES) barrier. 
Neither overhangs nor desorption are allowed;
see Fig.~\ref{fig:sim_defs} for a compact overview of these definitions. 
In the one-component case, 
the model is characterized by the four parameters $\epsilon=\eaa, \esub=\esubA, \EES=\EES_\text{11}, \Gamma = D/F$. 
Actual growth experiments of organic thin films are characterized by $\epsilon= -15 \ldots -10$,
and $\Gamma = 10^9 \ldots 10^{11}$ which is impossible to simulate. 
Nevertheless,  at lower energies and smaller diffusion-to-flux ratios $\Gamma$,
the model shows similar growth modes 
as seen experimentally. 
These are (a) island growth from the start when $\esub$ is low enough, 
(b) layer--by--layer growth (LBL) and 
(c) 3D growth of varying degree.
Matching the single-species growth modes is thus determining our choice of parameters.
For CuPc (species 1), 
3D growth is observed, which is initially LBL. 
C60 (species 2) exhibits strong island growth.
Both growth modes can be approximately modeled by the choices 
$\epsilon_{11}=-3.0$, $\esubA=-2.7$, $\EES_\text{11}=3.0$ (CuPc) and 
$\epsilon_{22}=-3.0$, $\esubB=-1.0$, $\EES_\text{22}=3.0$ (C60).
The substrate size is 200 $\times$ 200 sites.
We employed (hybrid) kinetic Monte Carlo (KMC) simulations, 
i.e. the simulation was divided into discrete time steps, 
during each of which either a new particle was inserted into the box 
or a move (see above) was attempted for an already existing particle. 
Insertion happens at a random position $(x,y)$ at a height of $h = \text{height}(x,y) + 1$
and $\text{height}(x,y)$ is the film height at the point $(x,y)$ before insertion.
For the other moves, 
the algorithm would first choose a random site $(x,y)$ 
at which at least one particle had already been deposited, 
and then try to move the topmost particle at this site to a random neighboring site. 
The move is accepted with a probability $p = \min \left\lbrace 1, \exp(- \Delta E)  \right\rbrace$.
If the particle has to climb or step down one layer, 
an additional acceptance probability of $\exp(-\EES)$ applies.
This hybrid KMC approach allows us to reduce the amount of bookkeeping 
necessary to track the change in rates due to the local environment of a particle. 
On the other hand, of course, this leads to moves being rejected and simulation time being wasted. 
Whether or not a move is accepted, 
the simulation time is still incremented by a timestep $\Delta t$ 
of variable length (Poisson distributed), 
since this leads to more accurate dynamics \cite{ruiz_barlett_2009}.
The interaction energy $\epsilon_{12}=\epsilon_{21}$ 
and the Ehrlich--Schw\"obel barrier $\EES_{12}=\EES_{21}$ 
are cross-species parameters yet to be fixed. 
CuPc and C60 are de-mixing. 
For the two-component lattice gas in equilibrium,
$2\epsilon_{12}-(\epsilon_{11}+\epsilon_{22}) > 0.88$ leads to de-mixing, 
but the deposition dynamics requires larger values,
otherwise de-mixing is kinetically suppressed. 
We choose $\epsilon_{12} =-0.5$. 
For the cross-species Ehrlich-Schw\"obel barrier one might speculate 
that $\EES_{12}<\EES_{11[22]}$ 
since the introduction of a second species 
opens more kinetic pathways for a molecule 
to climb or step down a layer.
The presence of lower diffusion barriers facilitates the interlayer diffusion 
such that $\EES_{12}<\EES_{11[22]}$ is justified.
Here, we simply choose $\EES_{12}=0$.
Since $D \propto \kt\,e^{-E_\text{D}/(\kt)}$, 
the main control parameter in the model to study the effect of temperature variation is the ratio $\Gamma=D/F$
between diffusion $D$ and flux $F$. 
We investigated the values $\Gamma=10^4$ and $10^5$. 
The change in $\Gamma$ by a factor of 10 (with $F=\text{const.}$) 
corresponds to the experimental change in temperature 
from $T_1=310$ K to $T_2=400$ K 
if the diffusion barrier $E_\text{D} \approx 11 \,k_\text{B}T_1$,
which is a reasonable value.
The roughness $\sigma$ of a growing film 
was evaluated in the same way as it was described in the experimental section
as $\sigma=\sqrt{ \langle h^2 \rangle - \langle h \rangle^2 }$ 
with $h\equiv h(x,y)$ as the instantaneous height of the film 
at the lateral lattice position $(x,y)$.
Finally, 
we note that the simulations were stopped 
immediately after a predetermined amount of material had been deposited, 
without time for the structures to equilibrate. 
This is in contrast to AFM measurements of grown films, 
which are often performed several days after growth, 
allowing the structure to potentially undergo changes.

% simulation definitions

\clearpage

\section{Comparison and Discussion of Results}
\label{sec:results}

\subsection{Evolution of Roughness}
\label{sec:results_roughness}

\begin{figure}[tbp]
		\centering
		\includegraphics[width=0.7\columnwidth]{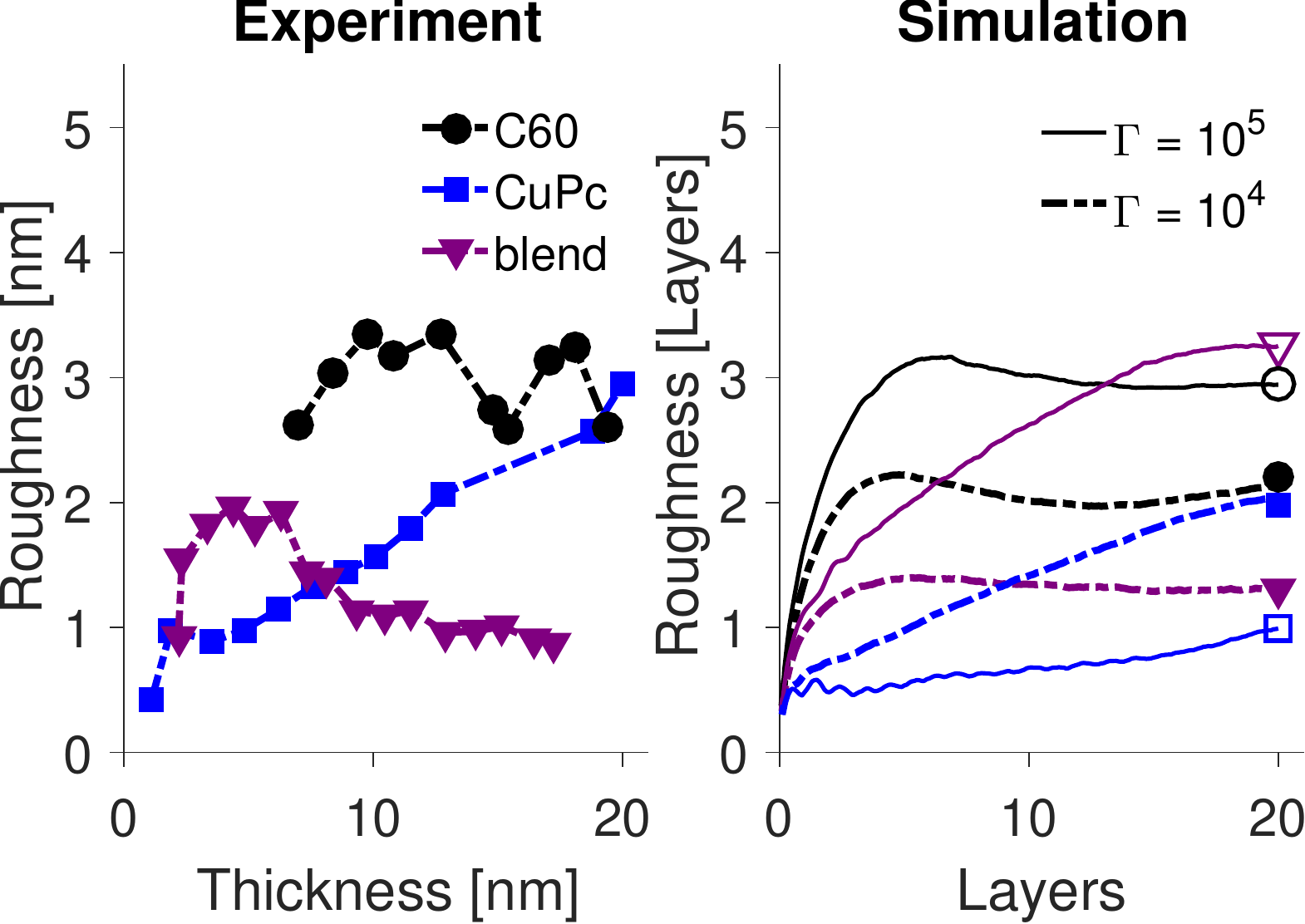}
		\caption{Evolution of the root mean square roughness determined from the Kiessig oscillations in XRR 
			at 310\,K (left)  and from simulations (right): 
			Dotted lines for $\Gamma=10^4$ (lower~$T$) and 
			solid lines for $\Gamma=10^5$ (higher~$T$). 
			Blue squares correspond to pure CuPc, black circles to pure C60, and purple triangles to the 1:1-blend.
			}
		\label{fig:roughness_evolution}
\end{figure}

\begin{figure}[tbp]
		\centering
		\includegraphics[width=0.7\columnwidth]{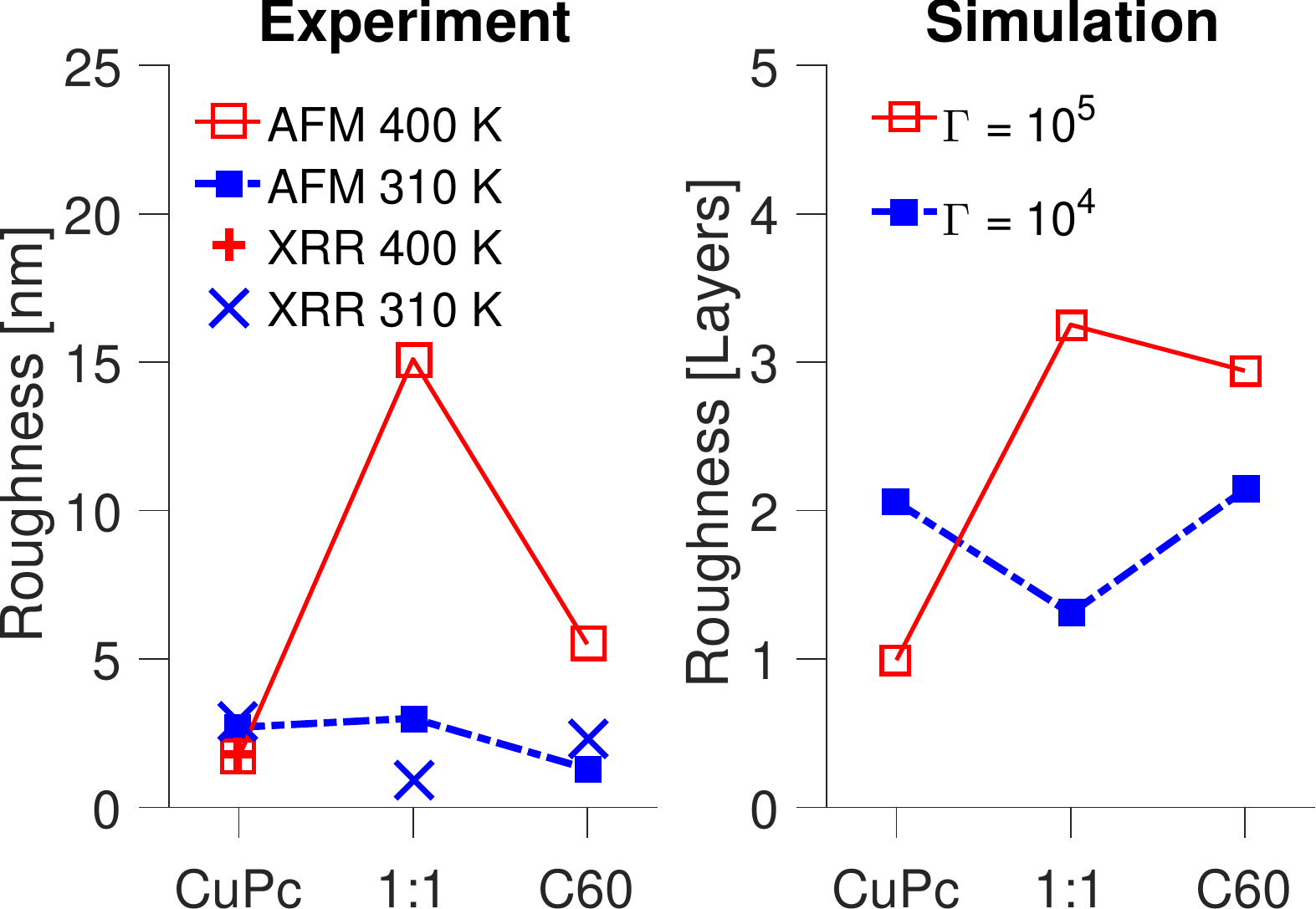}
		\caption{Post-growth root mean square roughness $\sigma$ determined from AFM images, XRR, and simulated height profiles
		at two different substrate temperatures: 310\,K ($\Gamma = 10^4$) and 400\,K ($\Gamma = 10^5$).}
		\label{fig:final_roughness}
\end{figure}

\noindent
Figure~\ref{fig:roughness_evolution} shows 
the evolution of the roughness $\sigma$
in the thin films grown at 310\,K
in comparison to the evolution of $\sigma$ 
simulated at $\Gamma = 10^4$ (dotted lines) 
and $\Gamma = 10^5$ (solid lines).
The corresponding experimental data sets of the blend and pure C60 at 400\,K 
could not be evaluated due to lacking Kiessig oscillations, see Fig.~\ref{fig:realtimeXRR}.
Furthermore, 
the roughness at 310\,K 
can only be determined after the first Kiessig oscillations become visible,
whereas the simulated value of $\sigma$ is available from the beginning of growth.
The experimental roughness is given in nanometers
and the simulated roughness is given in units of layers.
Depending on the particle species and the crystallographic direction, 
the vertical distances between two layers vary between 0.7\,nm and 1.3\,nm.
Assuming that one layer corresponds on average to 1\,nm,
the computational and experimental results agree well, 
both qualitatively as well as quantitatively
due to a suitable choice of interaction energies.
%
% post-growth roughness
%
Figure~\ref{fig:final_roughness} shows the post-growth roughness 
determined from AFM images of 20\,nm thin films
compared to the post-growth roughness 
determined from the simulated height profiles 
after deposition of 20\,layers.
Where feasible, 
the roughness was also determined 
from post-growth XRR.
Contrary to the XRR measurements, 
which were acquired \textit{in situ} directly after the growth,
the AFM measurements were carried out \textit{ex situ} several days after the growth
making post-growth effects possible.
Figure~\ref{fig:roughness_evolution} shows that
the evolution of roughness follows three different growth modes,
initial layer-by-layer (LBL) growth followed by rapid roughening in the case of pure CuPc,
rapid roughening from the beginning of growth in the case of pure C60
and 3D growth of varying degree for the blended thin films.
In the following, these growth modes will be discussed in turn.

\noindent
\textbf{CuPc:}
%
% LBL growth (simulation)
%
An alternating increase and decrease of roughness
was observed during the simulated growth 
of the first few layers of species 1 (CuPc) 
at $\Gamma = 10^5$ (higher $T$)
indicating LBL growth.
%
% Rapid roughening
%
Afterwards, 
the roughness of pure CuPc 
increases monotonously 
with time and thickness.
%
% Influence of T_sub
%
At $\Gamma = 10^4$ (lower $T$), 
the roughening starts immediately without any oscillations 
and hence, no LBL growth occurs.
It seems that the thickness 
after which the transition 
from LBL growth 
to rapid roughening happens
can be increased by increasing the substrate temperature.
%
% No ideal LBL growth
%
In simulations of very large systems, 
ideal LBL growth does not occur.
Instead, layer $n$ starts to grow before layer $n-1$ beneath is completed,
see Ref.\,\cite{Assis_2015_JStatMech} 
for investigations of a similar one-component model. 
The onset point $n$ of roughening however depends on the diffusion to flux ratio $\Gamma$, 
the Ehrlich-Schw\"obel barrier $\EES$ 
and also on the system size.
Overall it is a complicated kinetic effect.
%
% LBL growth (experiment)
%
Comparing the simulated data to our experiments,
the monotonously growing roughness at larger thicknesses 
is confirmed.
For a detailed evaluation of LBL growth, 
the temporal resolution of this experimental data set is too low.
%
% Comparison to literature
%
Experimental evidence for initial LBL growth can be found in prior studies 
on the thin film deposition of 
pentacene (PEN) 
\cite{Kowarik_2007_ThinSolidFilms},
diindenoperylene (DIP) 
\cite{Kowarik_2009_EuropPhysJSpecialTopics,Lorch_2017_JChemPhys,Durr_2003_PhysRevLett,Zhang_2007_SurfSci} 
and fluorinated copper phthalocyanine (F$_{16}$-CuPc) 
\cite{Yang_2015_SciRep}.
%
% Compare post-growth XRR and AFM
%
In the present study, 
the post-growth roughness values determined from XRR and AFM 
agree well and deviate at most by 2\,{\AA}, see Fig.~\ref{fig:final_roughness}.
%
% Compare post-growth exp and sim
%
The simulations result in a roughness 
of about 1 layer at $\Gamma = 10^5$ (higher~$T$) 
and 2 layers at $\Gamma = 10^4$ (lower~$T$).
These values are in good quantitative agreement with the experiment 
considering a vertical layer spacing of 13\,{\AA}, 
which was previously determined by XRR.
%
% Smoothing effect of increasing T_sub
%
Notable is that the substrate temperature increase 
has a smoothing effect on the CuPc thin film,
both in the experiment and in the simulation.
Elevating the substrate temperature
accelerates the downward diffusion of CuPc molecules 
and leads to smoother thin films
\cite{Reisz_2020_JApplCryst}.
The columnar growth of elongated CuPc crystals
along the substrate supports the smoothing further
\cite{Reisz_2020_JApplCryst}.

\noindent
\textbf{C60:}
The roughness of the C60 thin film increases rapidly 
in the beginning due to fast island formation
and remains almost constant for the rest of film growth,
see Fig.~\ref{fig:roughness_evolution}.
The rapid increase of the roughness
was reproduced computationally by the comparably weak C60-substrate interaction
leading to the formation of islands.
A ``dynamic wetting transition''
between initial LBL growth and island growth 
upon variation of $\esub$
was found in the simulations
for given $\epsilon$ and $\Gamma$.
For the chosen C60--C60 interaction $\epsilon=\epsilon_{22}=-3$
and an Ehrlich--Schw\"obel barrier of $\EES = -3$, 
this transition occurs
when $\esub < -2.44$ (for $\Gamma=10^4$) 
and $\esub < -2.59$ (for $\Gamma=10^5$),
which is substantially smaller than the chosen C60--substrate interaction of $\esubB=-1.0$.
Without an Ehrlich--Schw\"obel barrier ($\EES=0)$,
the transition occurs at slightly weaker substrate potentials
($\esub < -2.31$ for $\Gamma=10^4$ and $\esub < -2.52$ for $\Gamma=10^5$).
A small decrease of roughness arises after the initial roughening,
both in the simulation and the experiment.
It presumably stems from the coalescence of neighboring C60 islands
leading to a slightly smoother film.
%
% Post-growth roughness
%
In contrast to CuPc,
the post-growth roughness of C60 thin films increases 
when the substrate is heated during growth.
It was shown that C60 molecules react to substrate heating 
by island formation due to a molecular upward diffusion
\cite{Bommel_2015_PhysStatusSolidiRRL},
which explains the enhanced roughness at 400\,K
found in the present study.
This effect is well reproduced by the simulation,
see Fig.~\ref{fig:final_roughness}.
The deviation between the roughness 
determined from \textit{ex-situ} AFM 
and from \textit{in-situ} XRR at 310\,K
amounts to 1\,nm.
Statistical variations and moderate post-growth effects may account for this deviation.
The XRR-roughness of pure C60 and the blended thin film grown at 400\,K 
are not available due to the missing Kiessig oscillations.
The corresponding AFM values for C60 and the blend at 400\,K 
are higher than the values simulated at $\Gamma = 10^5$, 
possibly because of stronger post-growth effects.

\noindent
\textbf{Blends:}
The roughness in blended CuPc-C60 thin films grown at 310\,K 
increases rapidly in the beginning, 
decreases after the first five nanometers
and falls considerably below the roughness of the 
pure component films (Fig.~\ref{fig:roughness_evolution}). 
This is a quite remarkable behavior but seems to be a more generic
feature of blends
\cite{Hinderhofer_2020}.
In our simulations,
the effect can be reproduced only with $\EES_{12}<\EES_{11[22]}$, 
i.e. smaller cross-species ES barrriers.
Through the analysis of simulation snapshots
at different stages of growth
(see Sec.~\ref{sec:Real_Space_Images}),
two competing processes can be observed at $\Gamma = 10^4$ (lower $T$).
The nucleation of C60 islands increases the roughness in the beginning of growth
while CuPc fills the gaps in between and smoothes out the corrugations.
With the C60 domains expanding laterally, 
the space for CuPc is getting smaller and smaller.
As a result, 
CuPc has to grow vertically 
until it reaches the height of C60 domains
such that its smoothing impact finally prevails.
However, 
the smoothing effect (as seen in the overall roughness) is less pronounced in the simulation 
compared to the experiment.
Anisotropic intermolecular interactions 
that are not included in the simulation model
may additionally contribute to the smoothing effect.
Another growth scenario can be observed at 400\,K.
The missing Kiessig oscillations 
indicate a rapid roughening 
which remains during the entire growth.
The more rapid roughening was confirmed by the simulation at $\Gamma = 10^5$.
The solid line in Fig.~\ref{fig:roughness_evolution} shows 
that the roughness of the blend is at first in between those of the pure species
before increasing and finally becoming higher than the roughness of the pure films.
The corresponding simulation snapshots in 
Sec.~\ref{sec:Real_Space_Images} explain this behavior 
by the tendency of particles 
to grow on top of domains of the respective other species,
which compensates the above mentioned smoothing effect
and makes the blended thin film even rougher.
%
% Post-growth roughness
%
As a result of this,
the post-growth roughness in Fig.~\ref{fig:final_roughness} shows that 
unlike the remarkably smooth blend grown at 310\,K,
the roughness at 400\,K exceeds
than the roughness of all other thin films
both in the experimental and in the simulated data.
The roughness determined from AFM is significantly larger than the roughness determined from the simulation
and it is five times larger than the value determined from \textit{in-situ} XRR,
which may be the result of a post-growth de-wetting effect.
The simulation ended when the deposition was finished
without time for post-growth effects
and the \textit{in-situ} XRR scans were carried out directly after the growth,
whereas the AFM images where acquired \textit{ex situ} several days after the growth.

% Evolution of Roughness

% Post-Growth Roughness

\subsection{Island Densities and Sizes}
\label{sec:island_densities}

\begin{figure}[tbp]
		\centering
		\includegraphics[width=0.7\columnwidth]{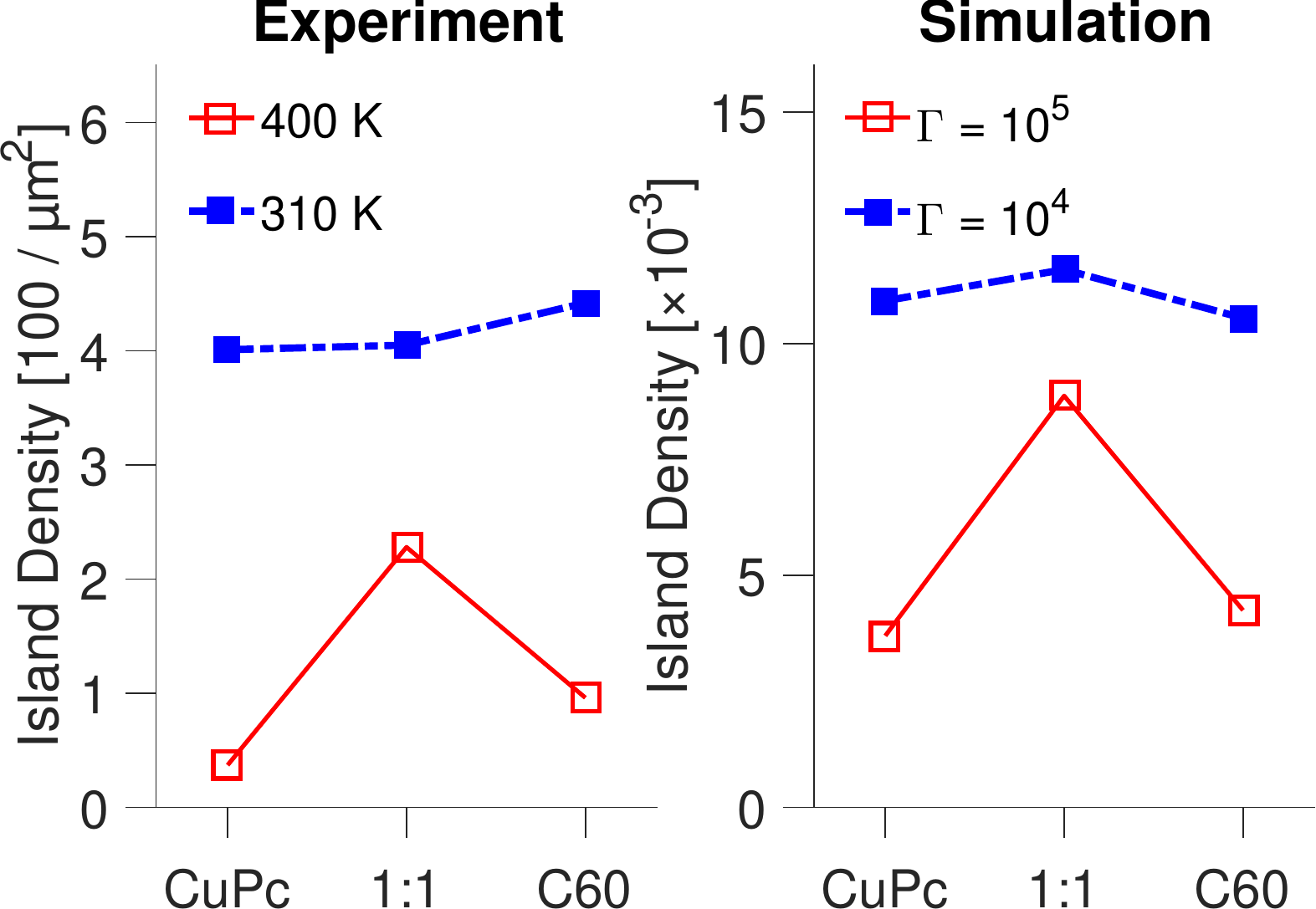}
		\caption{Island densities determined from AFM images and from simulated height profiles
		at two different substrate temperatures: 310\,K ($\Gamma = 10^4$) and 400\,K ($\Gamma = 10^5$).}
		\label{fig:island_densities}
\end{figure}

\noindent
Figure~\ref{fig:island_densities} presents the island densities 
extracted from the AFM images and the simulated height maps 
(see Sec.\ref{sec:Real_Space_Images}).
There are approximately four times less islands per area 
in the pure thin films grown at 400\,K than at 310\,K.
The island density at 400\,K was found to be below 100\,$\mum^{-2}$, 
whereas the number of islands reaches more than 400\,$\mum^{-2}$ at 310\,K
for both pure CuPc and pure C60 thin films.
Assuming a simple hexagonal distribution of islands, 
the inter-island distance $L$ would be 
$L = 1 / \sqrt{n \cdot \text{sin}(60^\circ)}$ 
with $n$ being the density of islands
and results in $L$ < 55\,nm at 310\,K 
and $L$ > 100\,nm at 400\,K.
As mentioned earlier,
the lateral sizes of coherently scattering domains determined from the width of GIXD peaks
are much smaller
indicating that each island consists of several crystallites with different orientations.
Regarding the blended thin films,
the size of coherently scattering domains goes down to less than 4\,nm.
However, there is no significant change of island densities
in blended thin films compared to the pure films grown at 310\,K.
The island density stays around 400\,$\mum^{-2}$ 
and is mainly determined by the numerous tiny islands 
surrounding a few very tall islands (see Sec.\,\ref{sec:Real_Space_Images}).
The tall islands reach heights of more than 50\,nm,
but their contribution to the roughness and to the island density is low 
because of their rare occurrence.
This behavior changes when raising the substrate temperature up to 400\,K.
The number of tall islands increases and their height reaches up to 100\,nm.
Regarding exclusively the density of tall islands which exceed a height of 50\,nm,
we find 1\,$\mum^{-2}$ at 310\,K and approximately 95\,$\mum^{-2}$ at 400\,K.
The overall island density at 400\,K is significantly larger than the island densities 
of the pure thin films grown at this temperature,
both in the experiment and in the simulation.
There is a good qualitative agreement between simulations and experiments.
Solely, the absolute values differ.
The experimentally observed tall islands at 310\,K with their large mutual distances
cannot be seen in simulations due to the limited substrate size of 200\,$\times$\,200 lattice sites.
Assuming that one lattice site corresponds to roughly 1\,nm$^2$, 
the simulated height maps represent only 0.4\% of the $3 \times 3\, \mum^2$ AFM images.
The simulated island density of 0.009 per lattice site 
for the blend system at $\Gamma=10^5$ 
corresponds then to $9000\ \mum^{-2}$, 
which is almost two orders of magnitude higher.
This is, however, to be expected, 
since we estimate that $\Gamma$ in the simulations 
is approximately 4 to 6 orders of magnitude lower 
than in an actual experiment 
leading to more and smaller islands.
Importantly, the qualitative agreement including all trends is excellent,
just on a smaller scale.

% Island Densities

\clearpage

\subsection{Real-Space Images and Overall Morphology}
\label{sec:Real_Space_Images}

\begin{figure*}[htbp]
		\includegraphics[width=0.9\textwidth]{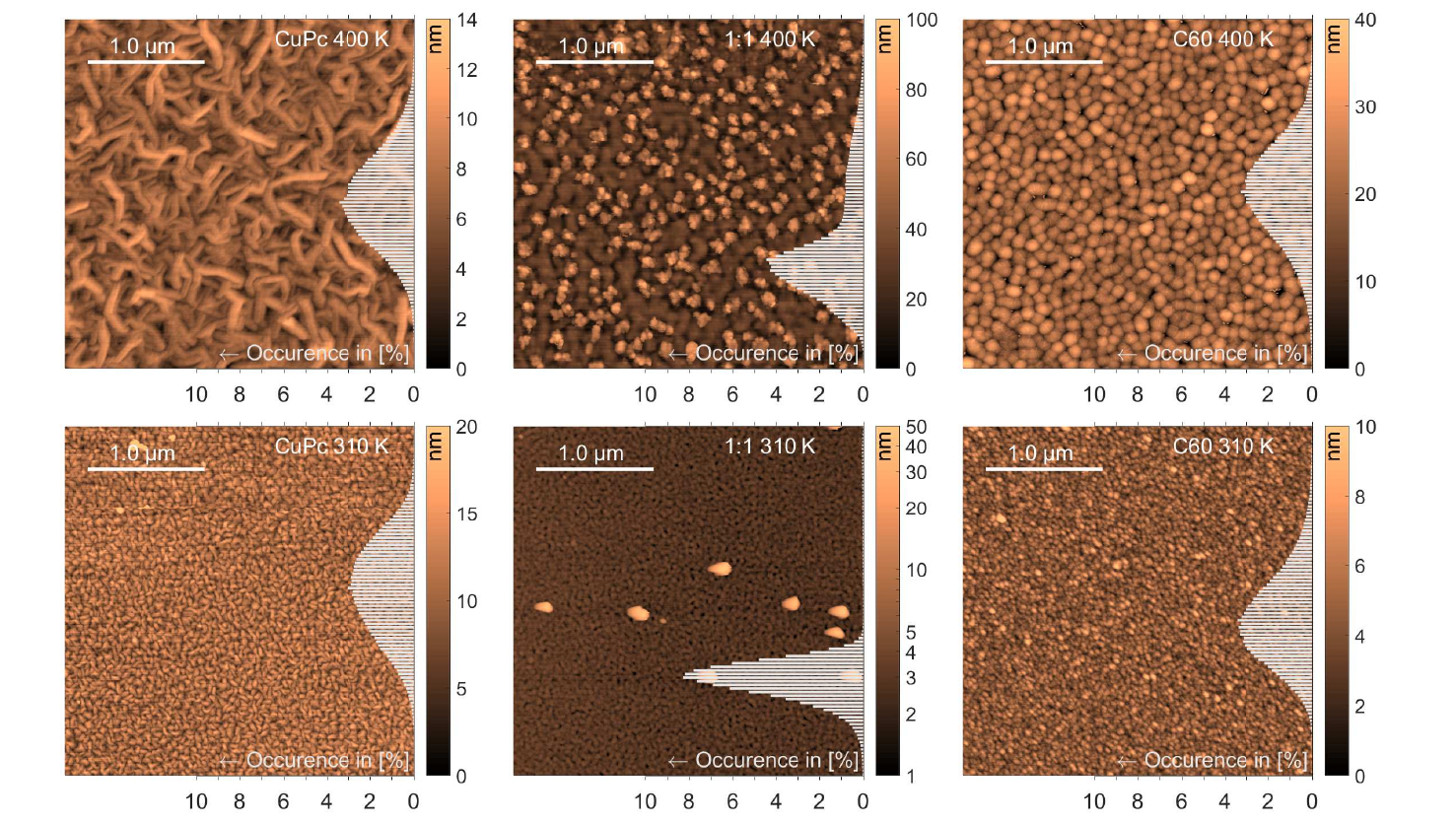}
		\caption{AFM images (3\,$\mu$m\,$\times$\,3\,$\mu$m) of pure CuPc (left), pure C60 (right) and the 1:1-blend (middle) 
			grown at 2\,{\AA}$/$min for 100\,min (average thickness $\approx$\,20\,nm) 
			at two different substrate temperatures 310\,K (lower row) and 400\,K (upper row).
			The distribution of heights are shown together with the color bar on the right side of each image.
			\vspace{-1em}}
		\label{fig:AFM_overview}
\end{figure*}

\begin{figure*}[htbp]
\includegraphics[width=0.9\textwidth]{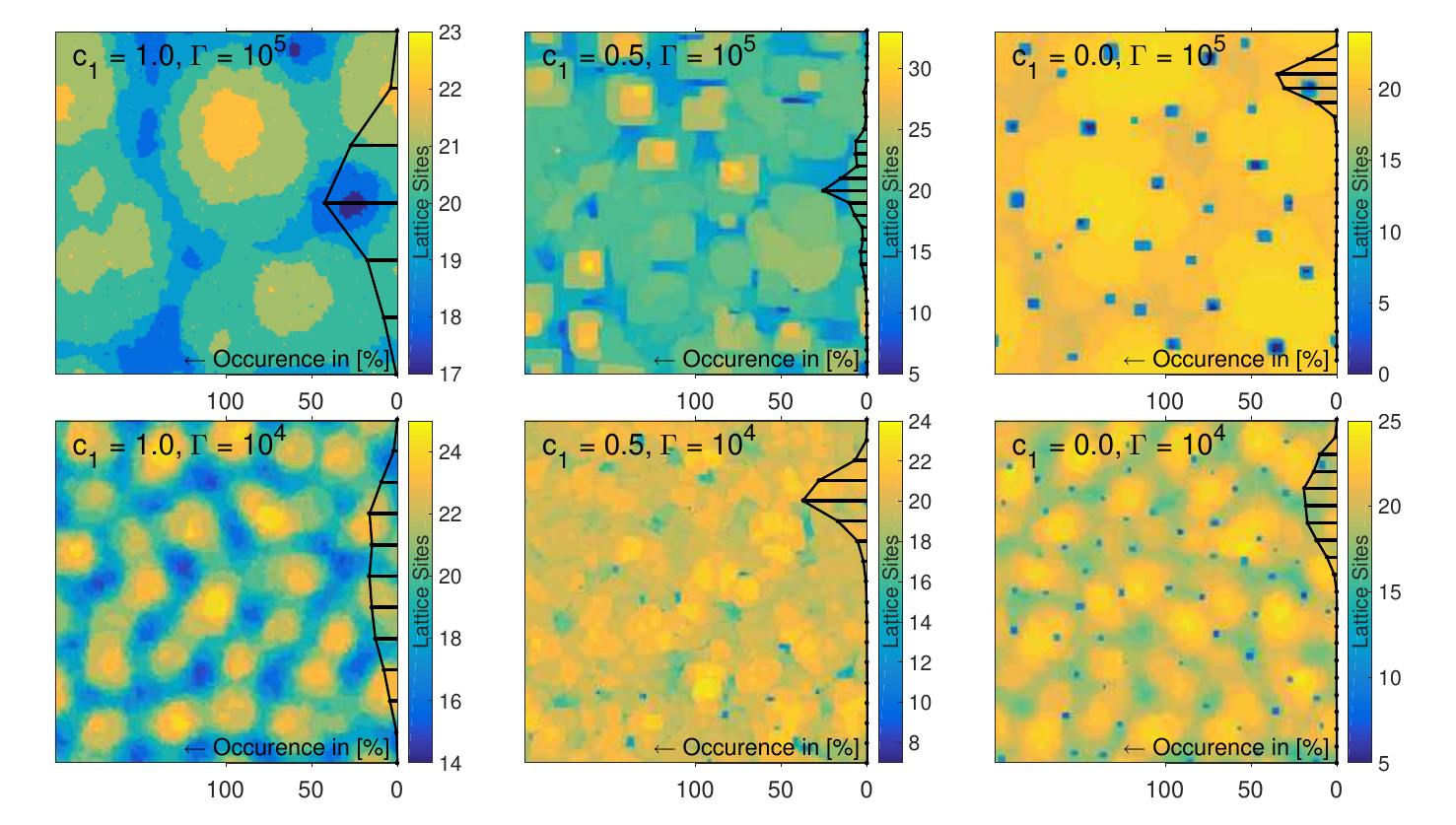}
\caption{Height maps of species~1 (CuPc, left), species~2 (C60, right) and the 1:1-blend (middle),
		in which the concentration of species 1 is $c_1 = 0.5$,
		simulated at different ratios $\Gamma = 10^4$ (lower~T, lower~row) and $\Gamma = 10^5$ (higher~T, upper~row).
		The total amount of particles deposited on the $200 \times 200$ lattice
		was $20 \times (200)^2$,
		which corresponds to effectively 20 completely filled layers ($\approx$\,20\,nm).
		\vspace{-2em}}
\label{fig:heights_simulation}
\end{figure*}

\begin{figure*}[htbp]
\includegraphics[width=\textwidth]{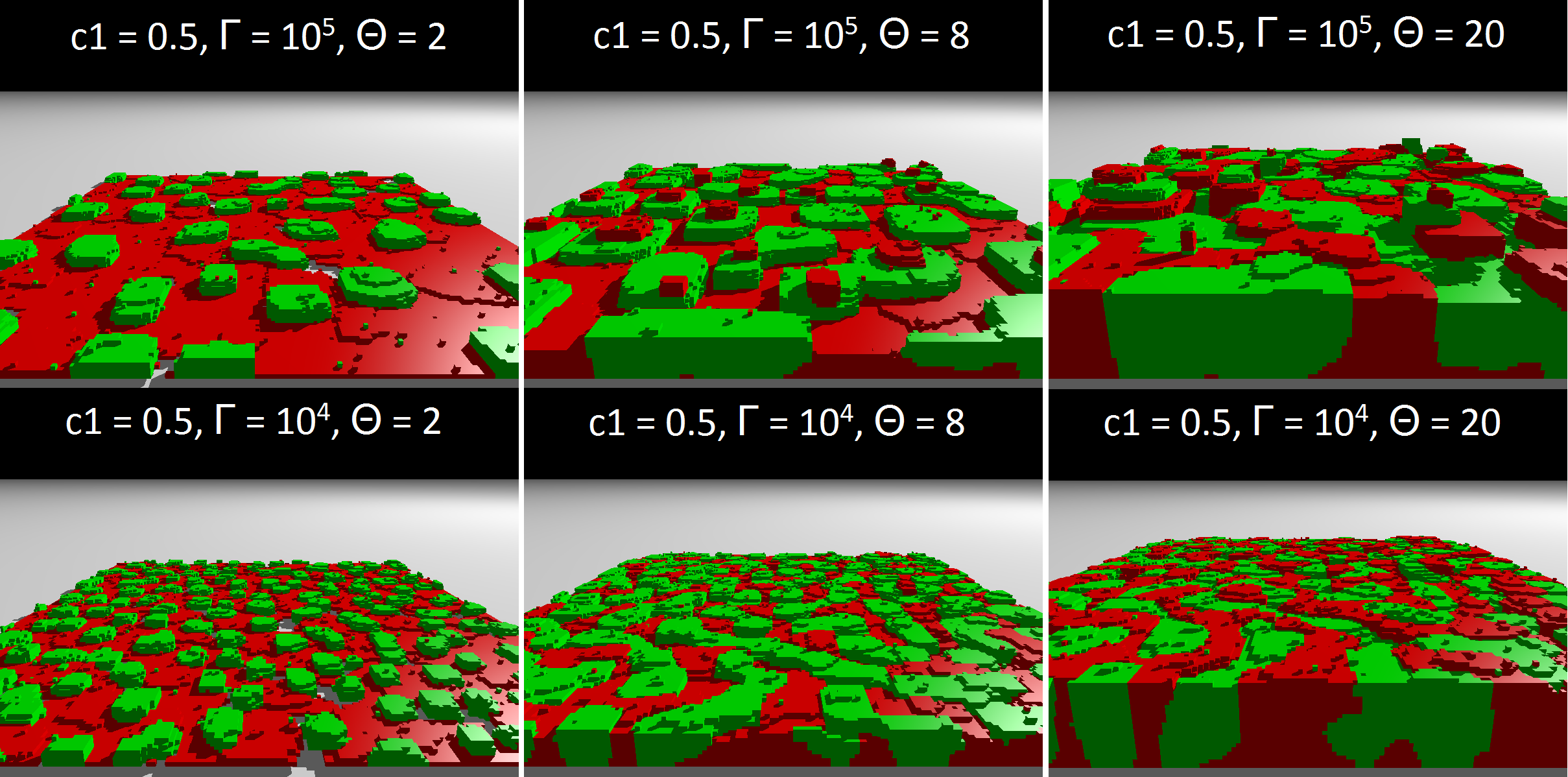}
\caption{Simulated snapshots on $200 \times 200$ lattice sites
		at a CuPc concentration of $c_1 = 0.5$ (red)
		and different ratios $\Gamma = 10^4$ and $\Gamma = 10^5$.
		The concentration of C60 is $c_2 = 1 - c_1$ (green).
		The total amount of deposited particles was $200 \times 200 \times \Theta$.}
\label{fig:snapshots}
\end{figure*}

\begin{figure}[tbp]
		\includegraphics[width=0.5\columnwidth]{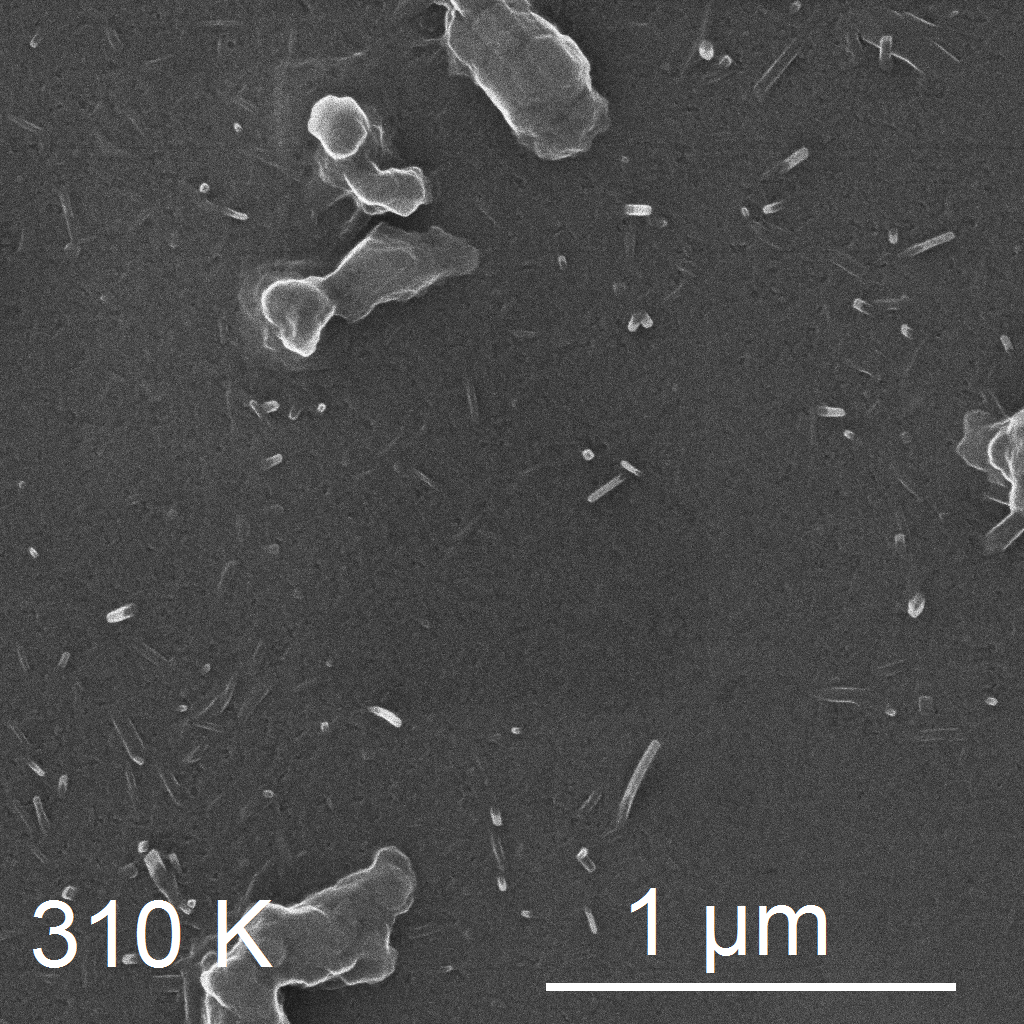}
		\caption{Helium ion microscopy (HIM) image
		of a CuPc-C60 blend (ratio 1:1) grown at 2\,{\AA}$/$min and a substrate temperature of 310\,K.
		The needle-like CuPc crystals protruding from the thin film are clearly recognizable.}
		\label{fig:HIM_images}
\end{figure}

\noindent
%
% AFM images
%
Typical AFM images are shown in Fig.~\ref{fig:AFM_overview}.
The distribution of heights is plotted on the right side of each image together with the color bar.
Note that the lowest point of the image is set to 0\,nm,
which is not necessarily the substrate.
There might be further completely filled layers of organic molecules below 0\,nm
and hence the maximum height in the color bar is not necessarily the film thickness.
As a result of the molecular structure,
C60 forms round islands
while the anisotropic CuPc molecules assemble in worm-like islands.
The increased islands size together with the reduced island density 
is clearly visible at 400\,K.
A faster diffusion of molecules can be assumed as reason 
for the increase of island sizes at elevated substrate temperatures.
A few tall islands arise in the mixed film at 310\,K
and their number increases substantially when going to 400\,K.
%
% Simulated height maps
%
The corresponding simulated height maps after deposition of 20 monolayers
are shown in Fig.~\ref{fig:heights_simulation}.
In a pure system of species 2 (C60), 
we see that there are ''holes'' in the film which reach in some cases down to the substrate
(see the right column). 
This is an artifact of the simulation method, 
since once the islands of the species merge, 
these holes occur due to the islands not being perfect squares. 
In order to fill the holes, 
new particles have to be deposited inside them, 
since they are too deep for surrounding particles to hop down and fill them up.
In contrast to species 2 (C60), 
we see how the pure growth of species 1 (CuPc)
leads to a very smooth film with no holes.
At higher $\Gamma$ we see a maximum deviation of $\pm 3$ MLs from the mean height
of this simulated thin film.
%
% Reference to the experiment
%
Although the lateral length scale in the simulations is much lower 
when compared to the AFM images,
there is qualitative agreement with the experiments
regarding the reduction of island densities upon increasing $T$ (or $\Gamma$).
Clearly, the simulations cannot capture the appearance of the worm-like CuPc islands 
or the rare tall islands in the mixed film. 
%
% Simulation Snapshots
%
Instead,
the simulations give a time-resolved picture of the 3D structure formation of the film. 
Example snapshots for the mixed film at the two values of $\Gamma$ 
and after deposition of 2, 8 and 20 monolayers are shown in Fig.~\ref{fig:snapshots}. 
In the following, the results from the analysis of the time-resolved 3D structure are discussed.
In the 1:1-blend, 
tall islands emerge on top of the film 
when increasing $\Gamma$ from $10^4$ to $10^5$, 
which are reminiscent of those seen in the experiments at 400\,K.
The formation process of these towers can be clearly seen in the snapshots: 
At first, species 2 (C60) will form islands on top of a wetting film of species 1 (CuPc). 
These islands then serve as nucleation points for clusters of species 1 which grow on top of them. 
On those, in turn, new clusters of species 2 will grow and this process may be continued. 
It appears that the absence of the cross-species ES-barrier ($\EES_{12}=0$) 
makes it favorable for the two species to grow clusters on top of clusters of the other species.
We recall that a low Ehrlich-Schw\"obel barrier for cross-species diffusion is reasonable
due to the larger number of possible kinetic pathways, 
which increases the probability for this kind of diffusion.
%
% HIM and SEM images
%
The simulation of the blended thin films gives rise to further questions to the experimental system.
How does the phase separation take place 
and which features belong to which molecule, 
i.e. to CuPc or to C60?
To answer this question, 
more insight is given by high resolution helium ion microscopy (HIM) images.
Needle-like CuPc crystallites protrude from the blended thin films,
see Fig.~\ref{fig:HIM_images} for example.
A previous study reported protruding needles 
in pure, vapor deposited CuPc films on silicon oxide
at larger film thicknesses of 50-500\,nm
\cite{Berger_2000_JMaterSci}.
It appears possible that  
the CuPc crystals are forced to grow vertically
if they are no longer exposed to the bare substrate, 
i.e. if they grow on already existing film layers.
Additional SEM images of blended thin films at different magnifications
are compiled in Fig.~\textcolor{blue}{S2} in the supporting information
\cite{Supporting_Info}.
At low magnifications,
those images provide an overview over a large area
and prove the homogeneous distribution of tall islands.
At high magnifications, 
the SEM images show needle-like crystals protruding from the thin films.
It is difficult to observe such thin needles by AFM.
Either the needles are bent downwards or they just break
when the AFM-tip moves across them.
In this case, 
the less invasive electron and ion scanning microscopy techniques are more suitable.
As discussed before,
the evolution of tall tower-like islands at $\Gamma = 10^5$ (higher $T$) 
can be explained by the ability of domains nucleating on top of the respective other species,
which was a result of the zero cross-species Ehrlich-Schw\"obel barrier.

% AFM images (2 A/min, 310 K and 400 K)

% Simulated Height Maps (Gamma = 1e4 and 1e5)

% Simulation Snapshots (Gamma = 1e4 and 1e5)

% HIM image and 3d-simulation

\clearpage

\section{Summary and Conclusion}
\label{sec:summary}

\noindent
This study provides a detailed quantitative picture of thin film growth of CuPc-C60 blends.
The experimental results are complemented by results from kinetic simulations of a generic, binary lattice model.
We regard the CuPc-C60 blend as a model for a phase-separating system and 
abstain from all peculiarities of molecular shape and anisotropic interactions in the simulations.
%
% Experiments
%
After initial layer-by-layer growth, rapid roughening was found in pure CuPc thin films,
whereas the C60 thin films started to roughen at a very early stage of growth due to a fast formation of islands.
It was shown that elevated substrate temperatures
lead to larger islands at four times lower island densities.
Furthermore, scanning electron and helium ion microscopy revealed needle-like islands
protruding from the blended thin films.
%
% Simulation
%
The simulations of the simple lattice model turned out to be well suited for understanding 
the influence of intermolecular interactions on the thin film growth
and they were capable of reproducing and rationalizing the behavior of overall quantities 
such as the evolution of roughness and the island densities.
An important principle for parameter selection lies in matching of the
pure-species growth modes even though the parameters for deposition rate and the interaction energies
were considerably off the experimental estimates. 
We performed simulations of the co-deposition of an island-forming particle species
and a quasi-layer-by-layer growing particle species, 
which can be seen as analogous to C60 and CuPc, respectively. 
The introduction of a species-dependent Ehrlich--Schw\"obel barrier led to films 
which were qualitatively similar to those observed in experiments.
% 
% increasing Gamma (T_sub)
%
By increasing the diffusion-to-deposition ratio $\Gamma = D/F$ by a factor of 10 
(which roughly corresponds to changing the substrate temperature from 310\,K to 400\,K), 
the following features were reproduced: 
(1) The blended film becomes rougher than the pure films.
(2) The island density in the blended film is significantly higher than in the pure films.
(3) Large islands (needles) start forming on top of the blended film.
Especially the last finding is of considerable interest,
since a possible mechanism 
for the formation of needle-like CuPc crystals 
protruding from the thin film
was found.
Large islands of one species nucleate on top of a cluster of the respective other species
leading to a pronounced roughness.
Simulations in which we used either no or a species-independent ES-barrier 
were unable to reproduce such a behavior 
indicating that this is a rationalization and model idealization of an effect which plays a role 
in actual film deposition experiments.

\begin{acknowledgments}

\noindent
We gratefully acknowledge the financial support of the German Research Foundation 
(Deutsche Forschungsgemeinschaft, DFG).
We thank the Paul Scherrer Institute 
for providing excellent facilities 
at the material science beamline MSX04SA of the Swiss Light Source
and the European Synchrotron Radiation Facility for excellent facilities
at the ID03 beamline.
We thank Ronny L\"offler and Markus Turad for providing the helium ion microscope at the LISA$+$ center in T\"ubingen.
G.~Duva gratefully acknowledges the Carl-Zeiss-Stiftung for support.
M.~Hodas further acknowledges the financial support of Alexander von Humboldt Foundation.

\end{acknowledgments}

%\bibliography{CuPc-C60-Refs}

\end{document}